\documentclass[english,12pt]{iopart}
\usepackage[T1]{fontenc}
\usepackage[latin1]{inputenc}
\usepackage{color}
\usepackage{babel}

\usepackage{graphicx}
\usepackage{amssymb}
\usepackage[unicode=true, pdfusetitle,
 bookmarks=true,bookmarksnumbered=false,bookmarksopen=false,
 breaklinks=true,pdfborder={0 0 0},backref=false,colorlinks=true]
 {hyperref}

\makeatletter
\usepackage{iopams}
\usepackage{setstack}

\usepackage{mathptmx}
\usepackage[amssymb]{SIunits}
\usepackage{cite}

\makeatletter

\renewcommand{\vec}{\mathbf}

\makeatother

\makeatother

\begin{document}

\title{Analytical model of non-Markovian decoherence in donor-based charge
quantum bits}

\author{F Lastra, S Reyes and S Wallentowitz}

\address{Facultad de Física, Pontificia Universidad Católica de Chile, Casilla
306, Santiago 22, Chile}
\begin{abstract}
We develop an analytical model for describing the dynamics of a donor-based
charge quantum bit (qubit). As a result, the quantum decoherence of
the qubit is analytically obtained and shown to reveal non-Markovian
features: The decoherence rate varies with time and even attains negative
values, generating a non-exponential decay of the electronic coherence
and a later recoherence. The resulting coherence time is inversely
proportional to the temperature, thus leading to low decoherence below
a material dependent characteristic temperature.
\end{abstract}


\maketitle

\section{Introduction}

Implementations of quantum bits (qubits) in solid-state systems represent
some of the most promising candidates for quantum-computation devices.
The constituent two-level systems have been experimentally realized
in a variety of systems, including quantum dots \cite{qudots}, impurities
in diamond \cite{impurities}, and superconductors \cite{superconductors}
among others. The implementation of solid-state quantum-information
devices has been object of increasing interest in the last decade
\cite{prokofev,amico,chirolli}.

Like other types of implementations, solid-state systems also suffer
from the inevitable interaction with their environment, which leads
to a loss of coherence of individual qubits and of entanglement between
them. Whereas the decoherence of many physical systems can be well
approximated by a Markovian process, a few prominent examples exist
where the coupling of the system with its environment reveals a memory
effect, leading to non-Markovian behavior and the possibility of recoherence
at later times. Among these are spin echo \cite{spin-echo} --- which
has been recently observed with phosphorus donors in silicon \cite{huebl},
photon echo \cite{photon-echo1,photon-echo2}, and collapse \cite{cummings}
and revival \cite{eberly,frahm,richter} in the spin-boson Jaynes--Cummings
interaction \cite{jaynes}. Collapse and revival may occur in various
physical systems, such as with atoms interacting with cavity electromagnetic
radiation \cite{haroche}, in the vibronic dynamics of electromagnetically
trapped ions \cite{blockley}, or in molecular vibration \cite{brif,wallentowitz}.
However, also other systems may show a non-Markovian dynamics in the
decay of their coherence. In particular, numerical calculations have
shown that donor-based charge qubits interacting with phonons reveal
a non-Markovian decay of electronic coherence \cite{thorwart}.

Donor-based charge qubits consist of a single electron shared by two
impurities in the semiconductor host. Besides being potentially scalable,
their manufacturing is largely facilitated by the technological base
provided by the existing semiconductor industry. One of the main sources
of decoherence in this system is the coupling of the electron to acoustic
phonons. This decoherence mechanism has been studied numerically for
$\milli\kelvin$ temperatures \cite{thorwart} and analytically for
$T=0\kelvin$ \cite{openov}. In the former reference, a non-exponential
decay of coherence has been numerically obtained, which indicates
a non-Markovian behavior of the electron-phonon coupling. On the other
hand, in Ref. \cite{openov} decoherence has been derived even for
zero temperature, where the bosonic phonon modes are in the vacuum
state. 

Non-Markovian coupling of a system to a reservoir, has received increasing
theoretical interest in recent years \cite{ivanov,ferrano,breuer2,anastopoulos,ferrano-1,cui,amin,chruscinski,xu,chen}.
An important conceptual issue is to obtain a solution when decay rates
attain negative values. In the literature, there exist numerical methods
to take into account negative decay rates. These comprise quantum
trajectory simulations in extended Hilbert spaces \cite{imamoglu,collet,diosi1,diosi2,breuer}
and alternative quantum-trajectory methods \cite{piilo,piilo2}. However,
for presently studied physical systems, negative decay rates have
been predicted only for extremely short time scales. 

We will show in this work that donor-based charge qubits, subject
to the interaction with acoustic phonons, do naturally show this feature
on time scales, that may be observable in experiment. Furthermore,
the model developed here contradicts the result of Ref. \cite{openov},
where decoherence is obtained even for zero temperature. From our
work, the coherence time is obtained as inversely proportional to
temperature, leading to vanishing decoherence at zero temperature.

\section{Model of Phonon Scattering with a Localized Electron\label{sec:Model-of-phonon}}

The Hamiltonian of the system under study is composed of three parts,
\begin{equation}
\hat{H}=\hat{H}_{{\rm e}}+\hat{H}_{{\rm ph}}+\hat{H}_{{\rm e-ph}},\end{equation}
where the electronic part is in principle composed by the Bloch electrons
plus the electrons bounded to the impurities. The quantized Fermionic
field for electrons with spin projection $\varsigma$ can be expanded
as\begin{equation}
\hat{\Psi}_{\varsigma}(\vec{r})=\sum_{n,\vec{q}}\hat{c}_{n,\vec{q},\varsigma}\psi_{n,\vec{q}}(\vec{r})+\sum_{a,n}\hat{d}_{a,n,\varsigma}\phi_{a,n}(\vec{r}),\label{eq:expansion}\end{equation}
where $\psi_{n,\vec{q}}(\vec{r})$ and $\phi_{a,n}(\vec{r})$ are
the Bloch functions of itinerant electrons and the wavefunction of
the electron localized at the donor site $a$, respectively. The operators
$\hat{c}_{n,\vec{q},\varsigma}$ and $\hat{d}_{a,n,\varsigma}$ annihilate
an electron with spin projection $\varsigma$ in the Bloch state $\psi_{n,\vec{q}}$
and in the localized state $\phi_{a,n}$, respectively. In our case
the impurities consist in pairs of donors (e.g. P) embedded in a semiconductor
substrate (e.g. Si). The full electronic Hamiltonian reads\begin{eqnarray}
\hat{H}_{{\rm e}} & = & \sum_{n,\vec{q},\varsigma}E_{n,\vec{q}}\hat{c}_{n,\vec{q},\varsigma}^{\dagger}\hat{c}_{n,\vec{q},\varsigma}+\sum_{a,n,\varsigma}\hbar\varepsilon_{a,n}\hat{d}_{a,n,\varsigma}^{\dagger}\hat{d}_{a,n,\varsigma}\nonumber \\
 &  & +\frac{1}{2}\sum_{n,n^{\prime}}\hbar\Delta_{n,n^{\prime}}\left(\hat{d}_{+,n}^{\dagger}\hat{d}_{-,n^{\prime}}+\hat{d}_{-,n^{\prime}}^{\dagger}\hat{d}_{+,n}\right),\label{eq:H-e}\end{eqnarray}
where $E_{n,\vec{q}}$ and $\hbar\varepsilon_{a,n}$ are the energies
of Bloch and localized electron, respectively. Here we added the third
term which is responsible for tunneling between localized states at
different donor sites with rates $\Delta_{n,n^{\prime}}$.

The phonon Hamiltonian corresponds to the energy of vibrational excitations
of the crystal lattice of the host semiconductor (Si). These excitations
are bosonic in nature and we can write, \begin{equation}
\hat{H}_{{\rm ph}}=\sum_{\vec{k},\sigma}\hbar\omega_{\vec{k},\sigma}\hat{b}_{\vec{k},\sigma}^{\dagger}\hat{b}_{\vec{k},\sigma},\label{eq:H-ph}\end{equation}
where the bosonic operator $\hat{b}_{\vec{k},\sigma}$ annihilates
a phonon with wavevector $\vec{k}$, polarization $\sigma$ ($\sigma=1,2,3$)
and energy $\hbar\omega_{\vec{k},\sigma}$. We will consider only
acoustic phonons, as the larger energies of optical phonons make them
practically unpopulated at room temperature and below.

\subsection{Electron-phonon interaction}

The electrons couple to the lattice vibrations through the third part
of the Hamiltonian, \begin{equation}
\hat{H}_{{\rm e-ph}}=e\int d^{3}r\hat{\rho}_{{\rm e}}(\vec{r})\left\{ \sum_{n}\hat{\vec{u}}_{n}\cdot\nabla V(\vec{r}-\vec{R}_{n})\right\} ,\label{eq:H-e-ph}\end{equation}
where the sum is performed over the lattice sites. Here $e$ is the
elementary electric charge and $\hat{\rho}_{{\rm e}}(\vec{r})$ is
the electron density. Furthermore, $V(\vec{r})$ is the Coulomb potential
of a single ion of the lattice with $\vec{R}_{n}$ and $\hat{\vec{u}}_{n}$
being the equilibrium position and quantized displacement of the ions,
respectively. 

In terms of the bosonic phonon operators we can write Eq. (\ref{eq:H-e-ph})
as\begin{equation}
\hat{H}_{{\rm e-ph}}=\frac{1}{\mathcal{V}}\int d^{3}r\hat{\rho}_{{\rm e}}(\vec{r})\sum_{\vec{k},\sigma}\sum_{\vec{K}}\kappa_{\vec{k},\vec{K},\sigma}\left(\hat{b}_{\vec{k},\sigma}+\hat{b}_{-\vec{k},\sigma}^{\dagger}\right)e^{i(\vec{k}+\vec{K})\cdot\vec{r}},\label{eq:H-e-ph2}\end{equation}
where $\mathcal{V}$ is the volume under consideration and the $\vec{k}$
and $\vec{K}$ sums are over the first Brillouin zone and the reciprocal
lattice, respectively. The coupling constant is derived as \begin{equation}
\kappa_{\vec{k},\vec{K},\sigma}=ie\sqrt{\frac{N\hbar}{2m_{0}\omega_{\vec{k},\sigma}}}\left(\vec{k}+\vec{K}\right)\cdot\vec{\epsilon}_{\vec{k},\sigma}V_{\vec{k}+\vec{K}}.\end{equation}
with $m_{0}$ and $N$ being the ionic mass in the unit cell and the
number of unit cells in the semiconductor crystal. Here the Fourier
coefficients of the ion potential are defined by\[
V(\vec{r})=\frac{1}{\mathcal{V}}\sum_{\vec{k}}V_{\vec{k}}e^{i\vec{k}\cdot\vec{r}}.\]

Using Eq. (\ref{eq:expansion}), the electron density can be written
as

\begin{equation}
\hat{\rho}_{{\rm e}}(\vec{r})=\hat{\rho}_{{\rm e}}^{({\rm B})}(\vec{r})+\hat{\rho}_{{\rm e}}^{({\rm BL})}(\vec{r})+\hat{\rho}_{{\rm e}}^{({\rm L})}(\vec{r}),\label{eq:density}\end{equation}
where the different terms correspond to pure Bloch electrons (B),
cross terms between Bloch and localized electrons (BL), and pure localized
electrons (L). Inserting Eq. (\ref{eq:density}) into (\ref{eq:H-e-ph2}),
the second term (BL) will produce phonon-induced transitions between
Bloch and localized electrons. However, for $T\leq300\kelvin$ the
thermal energy of phonons is not sufficient to drive such transitions.
Therefore, we can neglect the BL term together with the B term, since
under these circumstances the latter does not affect the dynamics
of the localized electron under concern. Thus, we are left with the
electron density \begin{equation}
\hat{\rho}_{{\rm e}}^{({\rm L})}(\vec{r})=\sum_{a,b}\sum_{n,n^{\prime}}\hat{d}_{a,n,\varsigma}^{\dagger}\hat{d}_{b,n^{\prime},\varsigma}\phi_{a,n}^{\ast}(\vec{r})\phi_{b,n^{\prime}}(\vec{r}),\label{eq:density2}\end{equation}

For sufficient high temperatures, where the average thermal phonon
energy, $k_{B}T$, is of the order of the electronic excitation energies
of individual donor sites, phonon-induced electronic transitions $n\to n^{\prime}$
($n\neq n^{\prime}$) within the individual donor sites ($a=b$) or
between the donor sites ($a\neq b$) may occur. In this case, resonant
phonon scattering occurs that leads to a decoherence mechanism with
a strong dependence on the inter-donor distance \cite{milburn-measure,fedorov}. 

In this work we consider a different temperature range, i.e. $k_{B}T$
is much smaller than the electronic excitation energies of individual
donor sites. In this case only the electronic ground states of the
donor sites can be populated and the density (\ref{eq:density2})
is further simplified to\begin{equation}
\hat{\rho}_{{\rm e}}^{({\rm L})}(\vec{r})=\sum_{a,b}\hat{d}_{a}^{\dagger}\hat{d}_{b}\phi_{a}^{\ast}(\vec{r})\phi_{b}(\vec{r}),\label{eq:density3}\end{equation}
where $\hat{d}_{a}=\hat{d}_{a,0,\varsigma}$ with $\varsigma$ being
the invariant spin projection of the single electron and $\phi_{a}(\vec{r})=\phi_{a,0}(\vec{r})$.
Inserting Eq. (\ref{eq:density3}) into the interaction Hamiltonian
(\ref{eq:H-e-ph2}) we obtain\begin{equation}
\hat{H}_{{\rm e-ph}}=\frac{1}{\mathcal{V}}\sum_{a,b=\pm}\sum_{\vec{k},\sigma}\sum_{\vec{K}}\kappa_{\vec{k},\vec{K},\sigma}F_{a,b}(\vec{k}+\vec{K})\hat{d}_{a}^{\dagger}\hat{d}_{b}\left(\hat{b}_{\vec{k},\sigma}+\hat{b}_{-\vec{k},\sigma}^{\dagger}\right),\label{eq:H-e-ph-approx1}\end{equation}
where the form factor is defined as\begin{equation}
F_{a,b}(\vec{k})=\int d^{3}r\phi_{a}^{*}(\vec{r})\phi_{b}(\vec{r})e^{i\vec{k}\cdot\vec{r}}.\label{form_factor}\end{equation}

Given that the rms spread of the localized electron wavefunction is
much larger than the periodicity of the semiconductor lattice, i.e.
it spreads over several lattice sites, the form factor (\ref{form_factor})
is non-vanishing only within the Brillouin zone, $F_{a,b}(\vec{k}+\vec{K})\approx0$
if $\vec{K}\neq\vec{0}$. Furthermore, given that the wavefunctions
of the two donor sites do not overlap, $F_{a,b}(\vec{k})=0$ for $a\neq b$.
Considering the medium as being approximately isotropic the interaction
Hamiltonian (\ref{eq:H-e-ph-approx1}) can then be simplified to\begin{equation}
\hat{H}_{{\rm e-ph}}=\frac{1}{\mathcal{V}}\sum_{a}\sum_{\vec{k}}\kappa_{k}F_{a}(\vec{k})\hat{d}_{a}^{\dagger}\hat{d}_{a}\left(\hat{b}_{\vec{k}}+\hat{b}_{-\vec{k}}^{\dagger}\right),\label{eq:H-e-ph-approx2}\end{equation}
where $F_{a}(\vec{k})=F_{a,a}(\vec{k})$ and where $\hat{b}_{\vec{k}}$
are annihilation operators of acoustic phonons with longitudinal polarization.
 The simplified coupling constant is\begin{equation}
\kappa_{k}=ie\sqrt{\frac{N\hbar}{2m_{0}\omega_{k}}}kV_{k}.\label{eq:kappa-def}\end{equation}
with linear dispersion $\omega_{k}=sk$, sound velocity $s$, and
with $V_{k}$ being the Fourier components of the ion potential.

\subsection{Electronic pseudo-spin}

In the basis of the localized-electron states $|a\rangle$ ($a=\pm$)
with $\langle\vec{r}|a\rangle=\phi_{a}(\vec{r})$ and for only a single
localized electron being present, i.e. $\hat{d}_{a}^{\dagger}\hat{d}_{a}|b\rangle=\delta_{a,b}|b\rangle$,
one may define the electronic pseudo spin-$\frac{1}{2}$ operator,
\begin{eqnarray}
\hat{S}_{x} & = & \frac{1}{2}\sum_{a\neq b}|a\rangle\langle b|,\\
\hat{S}_{y} & = & \frac{1}{2i}\sum_{a\neq b}a|a\rangle\langle b|,\\
\hat{S}_{z} & = & \frac{1}{2}\sum_{a}a|a\rangle\langle a|,\end{eqnarray}
and the electronic identity operator \[
\hat{I}_{{\rm e}}=\sum_{a}|a\rangle\langle a|.\]
With these definitions the sum over donor sites in Eq. (\ref{eq:H-e-ph-approx2})
is rewritten as\begin{equation}
\sum_{a}F_{a}(\vec{k})\hat{d}_{a}^{\dagger}\hat{d}_{a}=\hat{S}_{z}f_{\vec{k}}+\hat{I}_{{\rm e}}F_{\vec{k}},\label{eq:f-F}\end{equation}
with \begin{eqnarray}
F_{\vec{k}} & = & \frac{1}{2}\sum_{a}F_{a}(\vec{k}),\\
f_{\vec{k}} & = & F_{+}(\vec{k})-F_{-}(\vec{k}).\end{eqnarray}

The term proportional to the identity operator $\hat{I}_{{\rm e}}$
can be shown to lead to a phonon-induced shift of the electronic ground-state
energies $\hbar\varepsilon_{a,0}$ of the two donor sites ($a=\pm$),
see \ref{sec:Phonon-induced-shift-of}. It can be accounted for by
defining the displaced phonon operator \begin{equation}
\hat{a}_{\vec{k}}=\hat{b}_{\vec{k}}+\beta_{\vec{k}},\end{equation}
with the displacement being given in Eq. (\ref{eq:beta-def}) and
by redefinition of the electronic energies, see \ref{sec:Phonon-induced-shift-of}.
In this way the electronic Hamiltonian (\ref{eq:H-e}) can be simplified
to

\begin{equation}
\hat{H}_{{\rm e}}=\hbar\omega_{0}\hat{S}_{z}+\hbar\Delta\hat{S}_{x},\end{equation}
where $\omega_{0}$ is the transition frequency between the donor-site
electronic ground states, modified by the phonon-induced energy shift,
cf. Eq. (\ref{eq:omega0}) in \ref{sec:Phonon-induced-shift-of},
and $\Delta=\Delta_{0,0}$ is the tunneling rate between the ground
states of the donor sites. The interaction Hamiltonian (\ref{eq:H-e-ph-approx2}),
on the other hand, becomes\begin{equation}
\hat{H}_{{\rm e-ph}}=\sum_{\vec{k}}\hbar\hat{S}_{z}\left(g_{\vec{k}}\hat{a}_{\vec{k}}^{\dagger}+g_{\vec{k}}^{\ast}\hat{a}_{\vec{k}}\right),\end{equation}
where we defined the interaction rate\begin{equation}
g_{\vec{k}}=\frac{\kappa_{k}f_{-\vec{k}}}{\hbar\mathcal{V}}.\label{eq:interaction-rate}\end{equation}

\subsection{Interaction rate}

Assuming a Yukawa potential for the screened ions of charge $Ze$,
Eq. (\ref{eq:kappa-def}) becomes \begin{equation}
\kappa_{k}=i\frac{Ze^{2}}{\epsilon_{0}}\sqrt{\frac{N\hbar}{2m_{0}\omega_{k}}}\frac{k}{k^{2}+k_{s}^{2}},\end{equation}
with $k_{s}$ being the inverse screening length of the ionic potential.
For semiconductors (Si) and $T\leq300\kelvin$ the wavelengths of
populated phonon modes are much larger than the screening length,
i.e. $k\ll k_{s}$, so that we may approximate

\begin{equation}
\kappa_{k}\approx i\frac{Ze^{2}}{s\epsilon_{0}k_{s}^{2}}\sqrt{\frac{N\hbar\omega_{k}}{2m_{0}}}.\label{eq:kappa-approx}\end{equation}

The ground-state wavefunctions at the two donor sites are displaced
by the inter-donor distance vector $\vec{d}$, i.e. \begin{equation}
\phi_{\pm}(\vec{r})=\frac{1}{\sqrt{\pi R_{\pm}^{3}}}\exp\left[-\frac{|\vec{r}\mp\vec{d}/2|}{R_{\pm}}\right],\end{equation}
where $R_{a}$ ($a=\pm$) are the Bohr radii of the $s$-wave orbitals
of the two donor sites. Thus the form factors become\begin{equation}
F_{\pm}(\vec{k})=\frac{e^{\pm i\vec{k}\cdot\vec{d}/2}}{[1+(kR_{\pm}/2)^{2}]^{2}},\end{equation}
so that together with Eq. (\ref{eq:kappa-approx}) the interaction
rate (\ref{eq:interaction-rate}) becomes \begin{equation}
g_{\vec{k}}=\frac{D}{\hbar s}\sqrt{\frac{\hbar\omega_{k}}{2\rho_{m}\mathcal{V}}}\left\{ \frac{e^{-i\vec{k}\cdot\vec{d}/2}}{[1+(kR_{+}/2)^{2}]^{2}}-\frac{e^{i\vec{k}\cdot\vec{d}/2}}{[1+(kR_{-}/2)^{2}]^{2}}\right\} ,\end{equation}
where $\rho_{m}$ and $\mathcal{V}_{0}$ are the mass density and
unit-cell volume, respectively, and the deformation constant results
as $D=Ze^{2}/(\epsilon_{0}\mathcal{V}_{0}k_{s}^{2})$.

\section{Reduced Electronic Dynamics\label{sec:Electronic-Dynamics}}

We assume here that during the free evolution of the system, tunneling
is inhibited by either an applied potential barrier between the donor
sites, or by the different electronic ground-state energies of the
donor sites, $|\omega_{0}|\gg|\Delta|$, which can be provided for
by the application of a DC electric field. Under these circumstances
the Hamiltonian is given by\begin{equation}
\hat{H}=\hbar\omega_{0}\hat{S}_{z}+\sum_{\vec{k}}\hbar\omega_{k}\hat{a}_{\vec{k}}^{\dagger}\hat{a}_{\vec{k}}+\hbar\hat{S}_{z}\sum_{\vec{k}}\left(g_{\vec{k}}\hat{a}_{\vec{k}}^{\dagger}+g_{\vec{k}}^{\ast}\hat{a}_{\vec{k}}\right),\label{eq:H1}\end{equation}

\subsection{Electron-phonon eigenstates}

We look for the eigenstates of the Hamiltonian (\ref{eq:H1}),\begin{equation}
\hat{H}|E\rangle=E|E\rangle,\end{equation}
where from the Hamiltonian it is clear that they have the form\begin{equation}
|E\rangle\to|E_{m}\rangle=|m\rangle\otimes|\Phi_{m}\rangle,\end{equation}
with $\hat{S}_{z}|m\rangle=m|m\rangle$ and with $|\Phi_{m}\rangle$
being a solution of the phononic eigenvalue problem\begin{equation}
\hat{H}_{m}|\Phi_{m}\rangle=E_{m}|\Phi_{m}\rangle,\end{equation}
with the spin-projected phononic Hamiltonian being \begin{equation}
\hat{H}_{m}=\sum_{\vec{k}}\hbar\omega_{k}\hat{a}_{\vec{k}}^{\dagger}\hat{a}_{\vec{k}}+\hbar m\left[\omega_{0}+\sum_{\vec{k}}\left(g_{\vec{k}}\hat{a}_{\vec{k}}^{\dagger}+g_{\vec{k}}^{\ast}\hat{a}_{\vec{k}}\right)\right].\end{equation}

This Hamiltonian can be diagonalized as\begin{equation}
\hat{H}_{m}=\sum_{\vec{k}}\hbar\omega_{k}\hat{a}_{\vec{k},m}^{\dagger}\hat{a}_{\vec{k},m}+\hbar m\omega_{0},\end{equation}
where we used $m^{2}=1/4$ and discarded a constant energy term. Here
the operators \begin{equation}
\hat{a}_{\vec{k},m}=\hat{a}_{\vec{k}}+m\alpha_{\vec{k}},\end{equation}
are phonon operators with displacements conditioned on the electronic
state $m$. Their displacement amplitudes are given by\begin{equation}
\alpha_{\vec{k}}=\frac{g_{\vec{k}}}{\omega_{k}}.\end{equation}
The displacement of the phonon operators can be realized by the single-mode
displacement operators\begin{equation}
\hat{D}_{\vec{k}}(\alpha)=\exp\left(\alpha\hat{a}_{\vec{k}}^{\dagger}-\alpha^{\ast}\hat{a}_{\vec{k}}\right),\end{equation}
as\begin{equation}
\hat{a}_{\vec{k},m}=\hat{a}_{\vec{k}}+m\alpha_{k}=\hat{D}_{\vec{k}}^{\dagger}\left(m\alpha_{\vec{k}}\right)\hat{a}_{\vec{k}}\hat{D}_{\vec{k}}\left(m\alpha_{\vec{k}}\right).\end{equation}
Thus, the Hamiltonian can be written as\begin{equation}
\hat{H}_{m}=\hat{D}^{\dagger}\left(\left\{ m\alpha_{\vec{k}}\right\} \right)\hat{\tilde{H}}_{m}\hat{D}\left(\left\{ m\alpha_{\vec{k}}\right\} \right),\end{equation}
where \begin{equation}
\hat{\tilde{H}}_{m}=\sum_{\vec{k}}\hbar\omega_{k}\hat{a}_{\vec{k}}^{\dagger}\hat{a}_{\vec{k}}+\hbar m\omega_{0},\end{equation}
and where the multi-mode displacement operator is defined as\begin{equation}
\hat{D}(\{m\alpha_{k}\})=\prod_{\vec{k}}\hat{D}_{\vec{k}}(m\alpha_{\vec{k}})=\exp\left[\sum_{\vec{k}}m\left(\alpha_{\vec{k}}\hat{a}_{\vec{k}}^{\dagger}-\alpha_{\vec{k}}^{\ast}\hat{a}_{\vec{k}}\right)\right].\end{equation}

As a consequence, the phononic eigenvalue problem can be rewritten
as\begin{equation}
\hat{\tilde{H}}_{m}|\tilde{\Phi}_{m}\rangle=E_{m}|\tilde{\Phi}_{m}\rangle,\end{equation}
where the transformed eigenstates are\begin{equation}
|\tilde{\Phi}_{m}\rangle=\hat{D}\left(\left\{ m\alpha_{\vec{k}}\right\} \right)|\Phi_{m}\rangle.\end{equation}
Given that $\hat{\tilde{H}}_{m}$ is the Hamiltonian of an infinite
set of harmonic oscillators, these eigenstates are identified as those
of the harmonic oscillators, i.e. \begin{equation}
|\tilde{\Phi}_{m}\rangle\to|\{N_{\vec{k}}\}\rangle,\end{equation}
with $N_{\vec{k}}$ being the number of phonons in modes $\vec{k}$.
Consequently the energy spectrum is discrete and reads\begin{equation}
E_{m}\to E_{m,\{N_{\vec{k}}\}}=\sum_{\vec{k}}\hbar\omega_{k}N_{\vec{k}}+\hbar m\omega_{0}.\end{equation}
Transforming back to the original frame, the complete eigenstates
are therefore given by displaced number states:\begin{equation}
|E_{m,\{N_{\vec{k}}\}}\rangle=|m\rangle\otimes\hat{D}^{\dagger}\left(\left\{ m\alpha_{\vec{k}}\right\} \right)|\{N_{\vec{k}}\}\rangle.\label{eq:eigenstates}\end{equation}

The general solution of the reduced density operator of the spin system,
i.e. traced over the phonons, reads therefore\begin{eqnarray}
\hat{\varrho}_{S}(t) & = & \sum_{m,m^{\prime}}|m\rangle\langle m^{\prime}|\sum_{\{N_{\vec{k}}\}}\sum_{\{N_{\vec{k}}^{\prime}\}}\varrho_{m,\{N_{\lambda}\};m^{\prime},\{N_{\lambda}^{\prime}\}}\nonumber \\
 &  & \times\langle\{N_{\vec{k}}^{\prime}\}|\hat{D}\left(\left\{ m^{\prime}\alpha_{\vec{k}}\right\} \right)\hat{D}^{\dagger}\left(\left\{ m\alpha_{\vec{k}}\right\} \right)|\{N_{\vec{k}}\}\rangle\nonumber \\
 &  & \times\exp\left[-\frac{it}{\hbar}\left(E_{\{N_{\vec{k}}\},m}-E_{\{N_{\vec{k}}^{\prime}\},m^{\prime}}\right)\right].\label{eq:rho-reduced}\end{eqnarray}
where we used for the phononic trace ${\rm Tr}(|\psi\rangle\langle\phi|)=\langle\phi|\psi\rangle$
and where $\varrho_{m,\{N_{\vec{k}}\};m^{\prime},\{N_{\vec{k}}^{\prime}\}}$
is the initial density matrix of the complete spin-phonon system.
It can be easily seen, that the diagonal matrix elements of the solution
(\ref{eq:rho-reduced}) are constant, $\langle m|\hat{\varrho}_{S}(t)|m\rangle=\langle m|\hat{\varrho}_{S}(0)|m\rangle.$

We note that the initial reduced electronic density matrix elements
result from Eq. (\ref{eq:rho-reduced}) as \begin{eqnarray}
\langle m|\hat{\varrho}_{S}(0)|m\rangle & = & \sum_{\{N_{\vec{k}}\}}\varrho_{m,\{N_{\vec{k}}\};m,\{N_{\vec{k}}\}},\label{eq:rho-mm}\\
\langle g|\hat{\varrho}_{S}(0)|e\rangle & = & \sum_{\{N_{\vec{k}}\}}\sum_{\{N_{\vec{k}}^{\prime}\}}\varrho_{g,\{N_{\vec{k}}\};e,\{N_{\vec{k}}^{\prime}\}}f_{\{N_{\vec{k}}^{\prime}\},\{N_{\vec{k}}\}}\label{eq:rho-ge-1}\end{eqnarray}
and $\langle e|\hat{\varrho}_{S}(0)|g\rangle=\langle g|\hat{\varrho}_{S}(0)|e\rangle^{\ast}$.
The non-diagonal reduced matrix element (\ref{eq:rho-ge-1}) contains
the Franck--Condon transition amplitude \cite{franck,condon1,condon2},
well known from molecular physics,\begin{equation}
f_{\{N_{\vec{k}}^{\prime}\},\{N_{\vec{k}}\}}=\langle\{N_{\vec{k}}^{\prime}\}|\hat{D}(\{\alpha_{\vec{k}}\})|\{N_{\vec{k}}\}\rangle.\end{equation}
The presence of this factor is due to the fact, that the initial density
matrix of the complete spin-phonon system, $\varrho_{m,\{N_{\vec{k}}\};m^{\prime},\{N_{\vec{k}}^{\prime}\}}$,
is defined in the basis $|E_{m,\{N_{\vec{k}}\}}\rangle$, that differs
from the basis $|m\rangle\otimes|\{N_{\vec{k}}\}\rangle$ by the displacements
of the phonon state.

\subsection{Preparation of the initial electronic state}

The preparation of a coherent superposition of electronic states $|g\rangle$
and $|e\rangle$ must start from a well defined state, which we assume
to be the ground state $|g\rangle$. This can be experimentally realized
by applying a voltage $U\gg k_{{\rm B}}T$ between the donor sites.
When thermal equilibrium is reached after the switching on of the
voltage, the occupation probability of state $|g\rangle$ is very
close to one. The coherent electronic superposition can then be achieved
in two ways: Either by switching off the voltage to allow for coherent
tunneling between the donor sites, or by applying $\tera\hertz$ electromagnetic
radiation to perform a coherent Raman transition between states $|g\rangle$
and $|e\rangle$ \cite{openov2,koiller,tsukanov}. To observe the
decoherence of the prepared superposition state without the detrimental
effects of coherent transitions between the donor sites, the coherent
tunneling must be suppressed. This may be achieved either by applying
an electric field creating a potential barrier between the donor sites
or by switching on a sufficiently large voltage between the donor
sites.

As outlined above, the initial state $\hat{\varrho}(0)$ of the complete
system is the result of a state-preparation process in which the spin
originally starts in its ground state $|g\rangle$ and is coherently
transferred into a superposition of ground and excited states. In
the preparation process, a fraction $|\xi|^{2}$ of the population
is coherently transferred from the ground state $|g\rangle$ to the
excited state $|e\rangle$. However, as in the excited electronic
state each phonon mode receives a coherent displacement $\alpha_{\vec{k}}$,
cf. the eigenstates in Eq. (\ref{eq:eigenstates}), Franck--Condon
type transitions \cite{franck,condon1,condon2} occur for the phonons. 

The probability amplitude $\psi_{e,\{M_{\vec{k}}\}}$ to end in state
$|E_{e,\{M_{\vec{k}}\}}\rangle$ is given by\begin{eqnarray}
\psi_{e,\{M_{\vec{k}}\}} & = & \xi\sum_{\{N_{\vec{k}}\}}\langle E_{e,\{M_{\vec{k}}\}}|\hat{S}_{+}|E_{g,\{N_{\vec{k}}\}}\rangle\phi_{g,\{N_{\vec{k}}\}}\nonumber \\
 & = & \xi\sum_{\{N_{\vec{k}}\}}f_{\{M_{\vec{k}}\},\{N_{\vec{k}}\}}\phi_{g,\{N_{\vec{k}}\}},\label{eq:psi-e-def}\end{eqnarray}
where $\phi_{g,\{N_{\vec{k}}\}}$ are the probability amplitudes for
starting originally from states $|E_{g,\{N_{\vec{k}}\}}\rangle$.
On the other hand the probability amplitude $\psi_{g,\{N_{\vec{k}}\}}$
to stay in state $|E_{g,\{N_{\vec{k}}\}}\rangle$ is \begin{equation}
\psi_{g,\{N_{\vec{k}}\}}=\sqrt{1-|\xi|^{2}}\phi_{g,\{N_{\vec{k}}\}}.\label{eq:psi-g-def}\end{equation}
Thus the prepared complete initial state can be written as\begin{eqnarray}
|\psi(0)\rangle & = & \sum_{\{N_{\vec{k}}\}}\left(\psi_{g,\{N_{\vec{k}}\}}|E_{g,\{N_{\vec{k}}\}}\rangle+\psi_{e,\{N_{\vec{k}}\}}|E_{e,\{N_{\vec{k}}\}}\rangle\right),\end{eqnarray}
which by insertion of Eqs (\ref{eq:psi-e-def}) and (\ref{eq:psi-g-def})
becomes\begin{eqnarray}
|\psi(0)\rangle & = & \sum_{\{N_{\vec{k}}\}}\phi_{g,\{N_{\vec{k}}\}}\nonumber \\
 & \times & \left[\sqrt{1-|\xi|^{2}}|E_{g,\{N_{\vec{k}}\}}\rangle+\xi\hat{D}(\{\alpha_{\vec{k}}\})|E_{e,\{N_{\vec{k}}\}}\rangle\right]\end{eqnarray}
The corresponding initial density operator is $\hat{\varrho}(0)=|\psi(0)\rangle\langle\psi(0)|$
and reads\begin{eqnarray}
\hat{\varrho}(0) & = & \sum_{\{N_{\vec{k}}\}}\sum_{\{N_{\vec{k}}^{\prime}\}}\phi_{g,\{N_{\vec{k}}\}}\phi_{g,\{N_{\vec{k}}^{\prime}\}}^{\ast}\label{eq:rho0-1}\\
 & \times & \left\{ \left(1-|\xi|^{2}\right)|E_{g,\{N_{\vec{k}}\}}\rangle\langle E_{g,\{N_{\vec{k}}^{\prime}\}}|\right.\nonumber \\
 &  & +|\xi|^{2}\hat{D}(\{\alpha_{\vec{k}}\})|E_{e,\{N_{\vec{k}}\}}\rangle\langle E_{e,\{N_{\vec{k}}^{\prime}\}}|\hat{D}^{\dagger}(\{\alpha_{\vec{k}}\})\nonumber \\
 &  & +\xi^{\ast}\sqrt{1-|\xi|^{2}}|E_{g,\{N_{\vec{k}}\}}\rangle\langle E_{e,\{N_{\vec{k}}^{\prime}\}}|\hat{D}^{\dagger}(\{\alpha_{\vec{k}}\})\nonumber \\
 &  & \left.+\xi\sqrt{1-|\xi|^{2}}\hat{D}(\{\alpha_{\vec{k}}\})|E_{e,\{N_{\vec{k}}\}}\rangle\langle E_{g,\{N_{\vec{k}}^{\prime}\}}|\right\} \nonumber \end{eqnarray}

We assume that originally the spin-phonon system started from the
spin ground state and the phonons being in a thermal state, i.e.\begin{equation}
\hat{\varrho}=\sum_{\{N_{\vec{k}}\}}P_{\{N_{\vec{k}}\}}|E_{g,\{N_{\vec{k}}\}}\rangle\langle E_{g,\{N_{\vec{k}}\}}|,\end{equation}
with\begin{equation}
P_{\{N_{\vec{k}}\}}=Z^{-1}\exp\left(-\sum_{\vec{k}}\beta_{k}N_{\vec{k}}\right),\quad Z=\sum_{\{N_{\vec{k}}\}}\exp\left(-\sum_{\vec{k}}\beta_{k}N_{\vec{k}}\right),\end{equation}
where $\beta_{k}=\hbar\omega_{k}/(k_{{\rm B}}T)$. Thus we have to
substitute \begin{equation}
\phi_{g,\{N_{\vec{k}}\}}\phi_{g,\{N_{\vec{k}}^{\prime}\}}^{\ast}\to P_{\{N_{\vec{k}}\}}\prod_{\vec{k}}\delta_{N_{\vec{k}},N_{\vec{k}}^{\prime}},\end{equation}
by which the initial prepared spin-phonon density operator (\ref{eq:rho0-1})
becomes \begin{eqnarray}
\hat{\varrho}(0) & = & \sum_{\{N_{\vec{k}}\}}P_{\{N_{\vec{k}}\}}\\
 & \times & \left\{ \left(1-|\xi|^{2}\right)|E_{g,\{N_{\vec{k}}\}}\rangle\langle E_{g,\{N_{\vec{k}}\}}|\right.\nonumber \\
 &  & +|\xi|^{2}\hat{D}(\{\alpha_{\vec{k}}\})|E_{e,\{N_{\vec{k}}\}}\rangle\langle E_{e,\{N_{\vec{k}}\}}|\hat{D}^{\dagger}(\{\alpha_{\vec{k}}\})\nonumber \\
 &  & +\xi^{\ast}\sqrt{1-|\xi|^{2}}|E_{g,\{N_{\vec{k}}\}}\rangle\langle E_{e,\{N_{\vec{k}}\}}|\hat{D}^{\dagger}(\{\alpha_{\vec{k}}\})\nonumber \\
 &  & \left.+\xi\sqrt{1-|\xi|^{2}}\hat{D}(\{\alpha_{\vec{k}}\})|E_{e,\{N_{\vec{k}}\}}\rangle\langle E_{g,\{N_{\vec{k}}\}}|\right\} .\nonumber \end{eqnarray}

In the basis of the states (\ref{eq:eigenstates}) the corresponding
density matrix reads then\begin{eqnarray}
\varrho_{g,\{N_{\vec{k}}\};g,\{N_{\vec{k}}^{\prime}\}} & = & (1-|\xi|^{2})P_{\{N_{\vec{k}}\}}\prod_{\vec{k}}\delta_{N_{\vec{k}},N_{\vec{k}}^{\prime}},\\
\varrho_{e,\{N_{\vec{k}}\};e,\{N_{\vec{k}}^{\prime}\}} & = & |\xi|^{2}\sum_{\{M_{\vec{k}}\}}P_{\{M_{\vec{k}}\}}f_{\{N_{\vec{k}}\},\{M_{\vec{k}}\}}f_{\{N_{\vec{k}}^{\prime}\},\{M_{\vec{k}}\}}^{\ast},\\
\varrho_{g,\{N_{\vec{k}}\};e,\{N_{\vec{k}}^{\prime}\}} & = & \xi^{\ast}\sqrt{1-|\xi|^{2}}P_{\{N_{\vec{k}}\}}f_{\{N_{\vec{k}}^{\prime}\},\{N_{\vec{k}}\}}^{\ast}.\label{eq:rho-ge-initial}\end{eqnarray}
Furthermore, the reduced initial prepared spin density operator becomes\begin{eqnarray}
\hat{\varrho}_{S}(0) & = & (1-|\xi|^{2})|g\rangle\langle g|+|\xi|^{2}|e\rangle\langle e|\nonumber \\
 &  & +\xi^{\ast}\sqrt{1-|\xi|^{2}}|g\rangle\langle e|+\xi\sqrt{1-|\xi|^{2}}|e\rangle\langle g|,\end{eqnarray}
which corresponds to the perfectly pure superposition state\begin{equation}
|\psi_{S}(0)\rangle=\sqrt{1-|\xi|^{2}}|g\rangle+\xi|e\rangle.\end{equation}
Thus the presence of the phonons does not prevent a coherent preparation
of the initial spin state.

\subsection{Spin Decoherence}

From the general solution (\ref{eq:rho-reduced}), the time-dependent
off-diagonal matrix elements read \begin{eqnarray}
\langle g|\hat{\varrho}_{S}(t)|e\rangle & = & e^{i\omega_{0}t}\sum_{\{N_{\vec{k}}\}}\sum_{\{N_{\vec{k}}^{\prime}\}}\varrho_{g,\{N_{\vec{k}}\};e,\{N_{\vec{k}}^{\prime}\}}\label{eq:rho-ge}\\
 & \times & \langle\{N_{\vec{k}}^{\prime}\}|\hat{D}\left(\{\alpha_{\vec{k}}\}\right)|\{N_{\vec{k}}\}\rangle\exp\left[-i\sum_{\vec{k}}\omega_{k}(N_{\vec{k}}-N_{\vec{k}}^{\prime})t\right].\nonumber \end{eqnarray}
Inserting the initial state (\ref{eq:rho-ge-initial}) this matrix
element becomes\begin{eqnarray}
\langle g|\hat{\varrho}_{S}(t)|e\rangle & = & e^{i\omega_{0}t}\xi^{\ast}\sqrt{1-|\xi|^{2}}Z^{-1}\label{eq:rho-ge-time-dep}\\
 & \times & {\rm Tr}\left[\hat{D}^{\dagger}(\{\alpha_{\vec{k}}\})\hat{D}\left(\{\alpha_{\vec{k}}(t)\}\right)\exp\left(-\sum_{\vec{k}}\beta_{k}\hat{N}_{\vec{k}}\right)\right],\nonumber \end{eqnarray}
where \begin{equation}
\hat{D}\left(\{\alpha_{\vec{k}}(t)\}\right)=\exp\left(i\sum_{\vec{k}}\omega_{k}\hat{N}_{\vec{k}}t\right)\hat{D}\left(\{\alpha_{\vec{k}}\}\right)\exp\left(-i\sum_{\vec{k}}\omega_{k}\hat{N}_{\vec{k}}t\right)\end{equation}
with\begin{equation}
\alpha_{\vec{k}}(t)=\alpha_{\vec{k}}e^{-i\omega_{k}t}.\end{equation}

The displacements can be joined to obtain\begin{eqnarray}
\hat{D}^{\dagger}(\{\alpha_{\vec{k}}\})\hat{D}\left(\{\alpha_{\vec{k}}(t)\}\right) & = & \hat{D}(\{\alpha_{\vec{k}}(t)-\alpha_{\vec{k}}\})\nonumber \\
 & \times & \exp\left[-i\sum_{\vec{k}}|\alpha_{\vec{k}}|\sin(\omega_{k}t)\right],\end{eqnarray}
so that Eq. (\ref{eq:rho-ge-time-dep}) becomes\begin{eqnarray}
\langle g|\hat{\varrho}_{S}(t)|e\rangle & = & e^{i\omega_{0}t}\xi^{\ast}\sqrt{1-|\xi|^{2}}Z^{-1}\nonumber \\
 & \times & {\rm Tr}\left[\hat{D}(\{\alpha_{\vec{k}}(t)-\alpha_{\vec{k}}\})\exp\left(-\sum_{\vec{k}}\beta_{k}\hat{N}_{\vec{k}}\right)\right]\nonumber \\
 & \times & \exp\left[-i\sum_{\vec{k}}|\alpha_{\vec{k}}|\sin(\omega_{k}t)\right].\label{eq:rho_ge}\end{eqnarray}

The thermal averages can be calculated in phase space as\begin{equation}
{\rm Tr}\left[\hat{D}(\alpha)e^{-\beta\hat{N}}\right]=\frac{\int d^{2}\xi e^{\alpha\xi^{\ast}-\alpha^{\ast}\xi}e^{-\beta|\xi|^{2}}}{\int d^{2}\xi e^{-\beta|\xi|^{2}}}=e^{-|\alpha|^{2}/\beta}.\end{equation}
Thus Eq. (\ref{eq:rho-ge-1}) reads\begin{equation}
\langle g|\hat{\varrho}_{S}(t)|e\rangle=\langle g|\hat{\varrho}_{S}(0)|e\rangle\exp\left[i\omega_{0}t-i\phi(t)-\int_{0}^{t}d\tau\gamma(\tau)\right]\label{eq:ge-decay}\end{equation}
where the time-dependent phase reads\[
\phi(t)=\sum_{\vec{k}}|\alpha_{\vec{k}}|\sin(\omega_{k}t),\]
and the decoherence rate is given by\begin{eqnarray}
\gamma(t) & = & 2\sum_{\vec{k}}\frac{|\alpha_{\vec{k}}|^{2}\omega_{k}}{\beta_{k}}\sin(\omega_{k}t)\nonumber \\
 & = & \frac{2k_{{\rm B}}T}{\hbar}\sum_{\vec{k}}|\alpha_{\vec{k}}|^{2}\sin(\omega_{k}t).\end{eqnarray}

\subsection{Decoherence rate}

The mode sum is evaluated and after integration over the spherical
angles of the mode's wavevector, the rate becomes\begin{eqnarray}
 &  & \gamma(t)=\frac{k_{{\rm B}}TD^{2}}{(2\pi)^{2}s^{3}\hbar^{2}\rho_{m}}\int_{-\infty}^{\infty}dk\left\{ \sum_{a}\frac{k\sin(skt)}{\left[1+(kR_{a}/2)^{2}\right]^{4}}\right.\nonumber \\
 &  & \left.+\frac{\cos(|d+st|k)-\cos(|d-st|k)}{d\left[1+(kR_{-}/2)^{2}\right]^{2}\left[1+(kR_{+}/2)^{2}\right]^{2}}\right\} .\label{eq:gamma-eval1}\end{eqnarray}
 It can be written as\begin{equation}
\gamma(t)=\sum_{a\neq b}\gamma_{a,b}(t),\label{eq:def-gamma}\end{equation}
where we defined\begin{eqnarray}
\gamma_{a,b}(t) & = & \frac{\Gamma_{T}}{\eta_{a}^{2}}\left\{ \left[\frac{1}{3}\left(\frac{t}{\tau_{d}\eta_{a}}\right)^{3}+\frac{1}{2}\left(\frac{t}{\tau_{d}\eta_{a}}\right)^{2}+\frac{1}{4}\left(\frac{t}{\tau_{d}\eta_{a}}\right)\right]e^{-2t/(\tau_{d}\eta_{a})}\right.\nonumber \\
 &  & +\left(\frac{\eta_{a}^{2}}{\eta_{a}^{2}-\eta_{b}^{2}}\right)^{2}\left[\left|1+\frac{t}{\tau_{d}}\right|+\frac{\eta_{a}}{2}\frac{\eta_{a}^{2}-5\eta_{b}^{2}}{\eta_{a}^{2}-\eta_{b}^{2}}\right]e^{-2|1+t/\tau_{d}|/\eta_{a}}\nonumber \\
 &  & \left.-\left(\frac{\eta_{a}^{2}}{\eta_{a}^{2}-\eta_{b}^{2}}\right)^{2}\left[\left|1-\frac{t}{\tau_{d}}\right|+\frac{\eta_{a}}{2}\frac{\eta_{a}^{2}-5\eta_{b}^{2}}{\eta_{a}^{2}-\eta_{b}^{2}}\right]e^{-2|1-t/\tau_{d}|/\eta_{a}}\right\} ,\label{eq:gamma-ab}\end{eqnarray}
where we defined the relative Bohr radii \[
\eta_{a}=R_{a}/d,\qquad(a=\pm),\]
and the time needed for a phonon to travel between the donor sites,
$\tau_{d}=d/s$. The temperature dependent rate is defined by \[
\Gamma_{T}=\left(\frac{T}{T_{0}}\right)\omega_{d},\]
where $\omega_{d}=2\pi/\tau_{d}$ is the angular frequency of a phonon
with wavelength $\lambda=d$. The material dependent temperature scale
is defined for convenience as \begin{equation}
k_{{\rm B}}T_{0}=N_{d}m_{0}s^{2}\left(\frac{\hbar\omega_{d}}{D}\right)^{2},\label{eq:def-Tc}\end{equation}
where $N_{d}$ is the number of unit cells within the volume $d^{3}$.

\section{Discussion}

\subsection{Time-dependent decoherence rate}

\begin{figure}
\includegraphics[width=0.6\columnwidth]{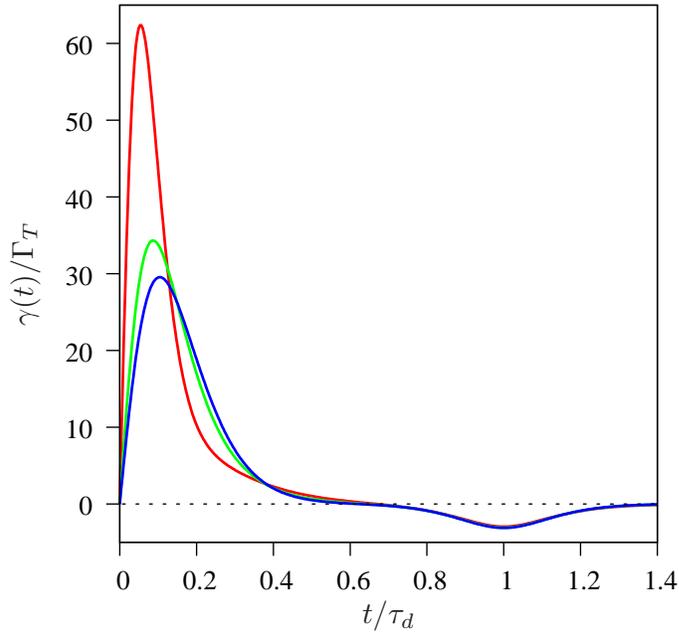}

\caption{\label{fig:Decoherence-rate-}Dimensionless decoherence rate $\gamma(t)/\Gamma_{T}$
as a function of the dimensionless time $t/\tau_{d}$ for $\eta=0.1$
and relative differences of the (relative) Bohr radii being: $\sigma=1.0$
(red curve), $\sigma=0.5$ (green curve), and $\sigma=0.1$ (blue
curve). Curves with $\sigma<0.1$ become indistinguishable from the
blue curve.}

\end{figure}

\begin{figure}
\includegraphics[width=0.6\columnwidth]{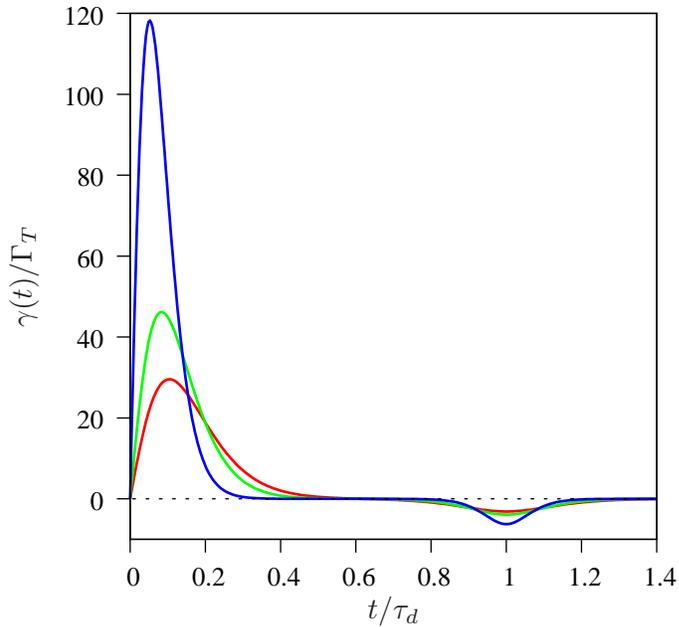}

\caption{\label{fig:Decoherence-rate-2}Dimensionless decoherence rate $\gamma(t)/\Gamma_{T}$
as a function of the dimensionless time $t/\tau_{d}$ for $\sigma=0.1$
and mean relative Bohr radius being: $\eta=0.1$ (red curve), $\eta=0.08$
(green curve), and $\eta=0.05$ (blue curve).}

\end{figure}

The initial value of the decoherence rate can be shown to be $\gamma(t=0)=0$.
Moreover, it is easily shown that it asymptotically decays to zero
for large times, $\gamma(t\to\infty)=0$. The dimensionless and temperature-independent
decoherence rate $\gamma(t)/\Gamma_{T}$ is plotted in Fig. \ref{fig:Decoherence-rate-}
as a function of the dimensionless time $t/\tau_{d}$. Here the mean
relative Bohr radius $\eta=\sum_{a}\eta_{a}/2$ has been chosen as
$\eta=0.1$, whereas the relative difference of (relative) Bohr radii,
$\sigma=\Delta\eta/\eta$ with $\Delta\eta=\eta_{+}-\eta_{-}$, is
varied for the three different curves. Curves for relative differences
of Bohr radii smaller than $0.1$ can be shown to converge to the
blue curve with $\sigma=0.1$. The positive and negative peaks become
more pronounced for decreasing ratio of Bohr radius over inter-site
distance, see Fig. \ref{fig:Decoherence-rate-2}. It is observed that
whereas the widths of the peaks decrease, their heights increase. 

It is readily recognized that the maximum decoherence rate always
occurs at the time $t\sim\tau_{d}\min(\eta_{a})$, which is the time
required for a phonon to cross the smallest donor site. A negative
peak of the decoherence rate, i.e. a maximal recoherence rate, is
observed always around the same time $t\sim\tau_{d}$. This corresponds
to the time needed for a phonon to travel between donor sites. Thus,
the timescales can be interpreted as the time needed for the phonon
to either travel within an impurity site or between the sites. In
the first case, when the phonon (as quasiparticle) is traveling within
one of the sites it destroys the inter-site coherence and does this
at maximum rate at the moment it has propagated a distance equal to
the smallest Bohr radius of the impurity states. In the latter case
the negative minimum, i.e. the recoherence peak, appears when the
phonon has traveled the distance between the impurities, thereby correlating
their states. 

The first time scale of the positive peak is produced by the first
term in Eq. (\ref{eq:gamma-ab}), whereas the second timescale of
the negative peak comes from the third term. The second term, on the
other hand, compensates the third one for times $t\ll\tau_{d}$ and
provides for $\gamma_{a,b}(t=0)=0$. It can be easily seen that when
expanding the decoherence rate (\ref{eq:def-gamma}) with respect
to the relative difference of (relative) Bohr radii $\sigma$ we obtain\[
\gamma(t)=\gamma_{0}(t)+\mathcal{O}\left(\sigma^{2}\right).\]
Here $\gamma_{0}(t)$ is the decoherence rate for the limiting case
of equal Bohr radii, $\sigma=0$, which is obtained as\begin{eqnarray*}
\fl\gamma_{0}(t) & = & \frac{\Gamma_{T}}{\eta^{2}}\Biggl\{\left[\frac{2}{3}\left(\frac{t/\tau_{d}}{\eta}\right)^{3}+\left(\frac{t/\tau_{d}}{\eta}\right)^{2}+\frac{1}{2}\left(\frac{t/\tau_{d}}{\eta}\right)\right]e^{-2(t/\tau_{d})/\eta}\\
\fl &  & +\eta\left[\frac{1}{6}\left(\frac{|1+t/\tau_{d}|}{\eta}\right)^{3}+\frac{1}{2}\left(\frac{|1+t/\tau_{d}|}{\eta}\right)^{2}+\frac{5}{8}\left(\frac{|1+t/\tau_{d}|}{\eta}\right)+\frac{5}{16}\right]e^{-2|1+t/\tau_{d}|/\eta}\\
\fl &  & -\eta\left[\frac{1}{6}\left(\frac{|1-t/\tau_{d}|}{\eta}\right)^{3}+\frac{1}{2}\left(\frac{|1-t/\tau_{d}|}{\eta}\right)^{2}+\frac{5}{8}\left(\frac{|1-t/\tau_{d}|}{\eta}\right)+\frac{5}{16}\right]e^{-2|1-t/\tau_{d}|/\eta}\Biggr\}.\end{eqnarray*}

The fact that the decoherence rate may attain negative values clearly
indicates a strong non-Markovian behavior. Thus, a certain memory
can be assigned to the phonon reservoir, with its timescale being
determined by the time needed for a phonon to travel the distance
between the impurities. The occurrence of negative decoherence rate
and non-Markovian behavior in an exactly solvable system, such as
presented here, is of high interest from the viewpoint of quantum-trajectory
theory. There, a controversy exists on how to find the correct trajectories
when decay rates may become negative \cite{breuer,collet,piilo,piilo2}.
However, up to now physically relevant models where such peculiar
rates can be observed at relevant timescales are largely unknown.
Thus, our model may be the first to show such features as experimentally
observable effect.

\subsection{Dependence on temperature}

The electronic coherence, i.e. the off-diagonal electronic density
matrix elements, decay with the function\[
g(t)=\exp\left[-\int_{0}^{t}dt^{\prime}\gamma(t^{\prime})\right],\]
cf. Eq. (\ref{eq:ge-decay}). It can be rewritten as\begin{equation}
g(t)=\left[G_{0}\left(t/\tau_{d}\right)\right]^{T/T_{0}},\label{eq:g(t)-result}\end{equation}
The decay function at $T=T_{0}$ with dimensionless time $x=t/\tau_{d}$
is obtained as\begin{equation}
G_{0}(x)=\exp\left[-\frac{2\pi}{\Gamma_{T}}\int_{0}^{x}dx^{\prime}\gamma(\tau_{d}x^{\prime})\right].\label{eq:G0-def}\end{equation}

Performing the time integral for the case $\sigma=0,$ i.e. for identical
donor sites, one obtains\begin{eqnarray}
\ln G_{0}(x) & = & -\left[\frac{1}{\eta}A\left(\frac{x}{\eta}\right)+B\left(\frac{1}{\eta}\right)-B\left(\frac{1+x}{\eta}\right)\right.\nonumber \\
 &  & +\Theta(1-x)\left[B\left(\frac{1}{\eta}\right)-B\left(\frac{1-x}{\eta}\right)\right]\nonumber \\
 &  & \left.+\left[1-\Theta(1-x)\right]\left[B\left(\frac{1}{\eta}\right)+B\left(\frac{x-1}{\eta}\right)-2B(0)\right]\right]\label{eq:G-equal}\end{eqnarray}
where $\Theta(x)$ is the Heaviside step function, whereas $A(x)$
and $B(x)$ describe the effects of the positive and negative peak
of the decoherence rate, respectively,\begin{eqnarray*}
A(x) & = & \pi\left[\frac{5}{4}-\left(\frac{2}{3}x^{3}+2x^{2}+\frac{5}{2}x+\frac{5}{4}\right)e^{-2x}\right],\\
B(x) & = & \pi\left(\frac{x^{3}}{6}+\frac{3}{4}x^{2}+\frac{11}{8}x+1\right)e^{-2x}.\end{eqnarray*}
Note, that the temperature-independent function (\ref{eq:G0-def})
depends in general on $\eta$ and $\sigma$, but it does not depend
on any parameter of the semiconductor material, such as sound speed
etc. This feature is provided for by the proper definition of the
critical temperature in Eq. (\ref{eq:def-Tc}) and the use of $t/\tau_{d}$
as dimensionless time in Eq. (\ref{eq:g(t)-result}).

In order to get a grasp on the experimental parameters involved, we
consider the donors to be phosphorus atoms, separated by a typical
distance of the order of $d\sim10\nano\meter$, embedded in a silicon
substrate. Silicon has a mass density $\rho_{m}=2.33\times10^{3}\kilogram\meter^{-3}$,
a deformation constant $D=8.6\electronvolt$, and sound speed $s=9\times10^{3}\meter\second^{-1}$.
With these values the time for a phonon to travel between the donor
sites is of the order of magnitude $\tau_{d}\sim1\pico\second$. Most
importantly, the characteristic temperature scale, as defined in Eq.
(\ref{eq:def-Tc}), is of the order of $T_{0}\sim10^{3}\kelvin$ for
this material. In consequence, all relevant operational temperatures
are much lower than this temperature scale ($T\ll T_{0}$).

\begin{figure}
\includegraphics[width=0.6\columnwidth]{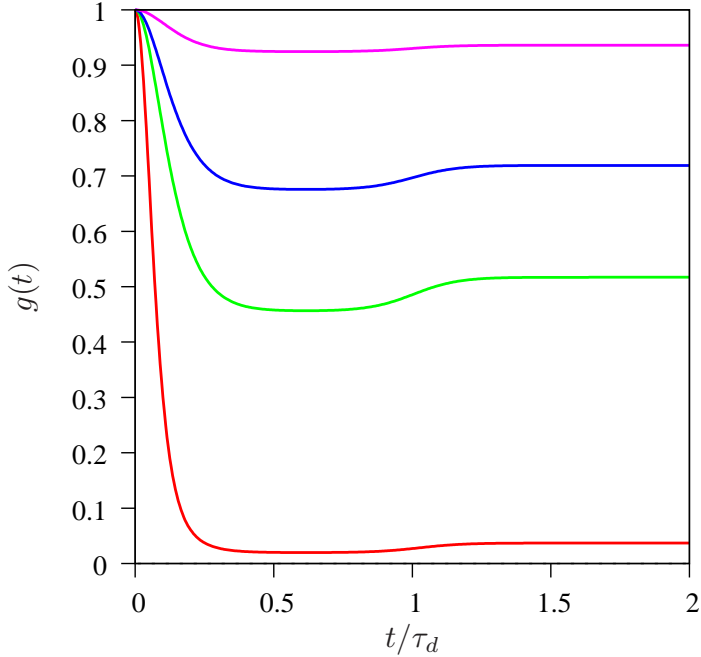}

\caption{\label{fig:Temperature-dependence-of}Temperature dependence of the
electronic decoherence function $g(t)$ as a function of dimensionless
time $t/\tau_{d}$ for $\eta=0.1$, $\sigma=0.0$ and temperatures:
$T/T_{0}=0.1$ (red curve), $T/T_{0}=0.02$ (green curve), $T/T_{0}=0.01$
(blue curve), and $T/T_{0}=0.002$ (purple curve).}

\end{figure}

In Fig. \ref{fig:Temperature-dependence-of} the decay function (\ref{eq:g(t)-result})
is shown for $\eta=0.1$ ($\sigma=0$) and various values of $T$
with $T\ll T_{0}$. It can be readily seen that apart from deviating
from the Markovian exponential decay it additionally is a non-monotonic
function, i.e. it first decays and then rises again due to the negative
decoherence rate at $t\sim\tau_{d}$. A simple definition of a coherence
time $\tau_{{\rm c}}$ by requiring $g(\tau_{{\rm c}})=1/e$, being
appropriate for an exponential decay, leads to problems of multi-valuedness
in this case. 

It is more convenient to specify the transient mean decoherence rate
$\bar{\gamma}$, corresponding to the rate averaged up to the time
$\tau_{d}$ when the recoherence starts,\[
\bar{\gamma}=\frac{1}{\tau_{d}}\int_{0}^{\tau_{d}}dt\gamma(t).\]
Employing the results in Eqs (\ref{eq:g(t)-result}) and (\ref{eq:G0-def}),
and using $\eta\tau_{d}=R/s$, the order of magnitude of the coherence
time of the system, $\tau_{c}=1/\bar{\gamma}$, is obtained for $\sigma=0$
and $\eta\ll1$ as\[
\tau_{c}\approx\frac{4}{5\pi}\left(\frac{R}{s}\right)\left(\frac{T_{0}}{T}\right).\]
Thus for $T=0\kelvin$ the coherence time becomes infinite and no
decoherence occurs%
\footnote{Obviously, other mechanisms, that were not considered here, may still
lead to decoherence at $T=0\kelvin$. %
}. This result is in contrast with recent results using a Wigner-Weisskopf
calculation but truncating the set of coupled differential equations,
where decoherence can be observed even for $T=0\kelvin$ \cite{openov}.

\section{Summary and Outlook}

In summary, we have studied the decoherence dynamics of a charge donor
quantum bit and have obtained results in agreement with numerical
studies \cite{thorwart} but in disagreement with the zero-temperature
theory proposed in Ref. \cite{openov}. Our treatment is entirely
analytic and exact and provides an analytical framework of a strongly
non-Markovian physical system for testing recently published quantum
trajectory methods for systems with negative decay rates \cite{breuer,breuer2,collet,piilo,piilo2}.

Our theory reveals that the dynamics of the electron-phonon system
is dominated by Franck--Condon type transitions that entangle electron
and phonon states by an electron conditioned coherent displacement
of the phonon modes. As a consequence, the reduced electron dynamics
shows two strong non-Markovian features. Firstly the initial decay
is not exponential but quadratic and secondly a later recoherence
is observed due to the occurrence of a negative decay rate. This recoherence
leads to a stationary and appreciable value of electronic coherence
for temperatures $T\ll T_{0}$, with $T_{0}\sim1000\kelvin$, which
thus should be observable at room temperature. Furthermore, as the
dynamics of the electronic coherence is described by an analytic function,
its dependence on temperature is obtained in analytic form as a simple
scaling law.

In conclusion, our results indicate that charge donor-based implementations
of quantum bits are promising candidates for quantum memory and computation
at feasible temperatures. However, further investigations are needed
to study other types of decoherence mechanisms that may affect this
type of system, such as coupling to nuclear spins or electromagnetic
fluctuations. Furthermore, the effects of coherent tunneling could
also be included in the future in a perturbative manner.

\ack{}{SW and FL acknowledge support by FONDECYT project no. 1095214.}

\appendix

\section{Phonon-induced shift of electronic energies\label{sec:Phonon-induced-shift-of}}

After inserting Eq. (\ref{eq:f-F}) into the Hamiltonian (\ref{eq:H-e-ph-approx2})
and using Eq. (\ref{eq:H-ph}), one obtains the complete Hamiltonian
as

\begin{equation}
\hat{H}=\hat{H}_{{\rm e}}+\sum_{\vec{k}}\hbar\omega_{k}\hat{b}_{\vec{k}}^{\dagger}\hat{b}_{\vec{k}}+\frac{1}{\mathcal{V}}\sum_{\vec{k}}\kappa_{k}\left(\hat{S}_{z}f(\vec{k})+\hat{N}_{{\rm e}}F(\vec{k})\right)\left(\hat{b}_{\vec{k}}+\hat{b}_{-\vec{k}}^{\dagger}\right).\end{equation}
Given only one localized electron, $\hat{N}_{{\rm e}}\to\hat{I}$
with $\hat{I}$ being the (electronic) identity operator, so that
the phonon part can be diagonalized by the use of the shifted phonon
operator \begin{equation}
\hat{a}_{\vec{k}}=\hat{b}_{\vec{k}}+\beta_{\vec{k}},\label{eq:a-def}\end{equation}
with the shift being \begin{equation}
\beta_{\vec{k}}=\frac{\kappa_{k}F^{\ast}(\vec{k})}{\mathcal{V}\hbar\omega_{k}}.\label{eq:beta-def}\end{equation}

Performing the approximations as described in Sec. \ref{sec:Model-of-phonon}
the electronic Hamiltonian reduces to \begin{equation}
\hat{H}_{{\rm e}}=\hbar\left(\varepsilon_{+,0}-\varepsilon_{-,0}\right)\hat{S}_{z}.\end{equation}
Due to the use of the displaced phonon operator (\ref{eq:a-def}),
this electronic Hamiltonian is modified to obtain the complete Hamiltonian
as\begin{equation}
\hat{H}_{{\rm e-ph}}=\sum_{\vec{k}}\hbar\omega_{k}\hat{a}_{\vec{k}}^{\dagger}\hat{a}_{\vec{k}}+\hbar\omega_{0}\hat{S}_{z}+\frac{1}{\mathcal{V}}\sum_{\vec{k}}\kappa_{k}f(\vec{k})\hat{S}_{z}\left(\hat{a}_{\vec{k}}+\hat{a}_{-\vec{k}}^{\dagger}\right),\end{equation}
where the shifted electronic transition frequency is\begin{equation}
\omega_{0}=\left(\varepsilon_{+,0}-\varepsilon_{-,0}\right)-\frac{1}{\hbar\mathcal{V}}\sum_{\vec{k}}\kappa_{k}f(\vec{k})\left(\beta_{\vec{k}}+\beta_{-\vec{k}}^{\ast}\right).\label{eq:omega0}\end{equation}

\end{document}